\documentclass[a4paper,12pt]{article}
\usepackage{amsmath,amsfonts,amssymb,amsthm,amstext,amscd}
\usepackage{latexsym}
\usepackage{hyperref}
\usepackage{graphicx}
\usepackage{caption}
\usepackage{comment}
\usepackage{graphicx}
\usepackage{cite}
\usepackage{color}
\usepackage{setspace}
\usepackage[mathscr]{eucal}
\usepackage{xcolor}
\usepackage{titletoc}

{%
	\definecolor{BLACK}{gray}{0}
	\definecolor{WHITE}{gray}{1}
	\definecolor{RED}{rgb}{1,0,0}
	\definecolor{GREEN}{rgb}{0,1,0}
	\definecolor{BLUE}{rgb}{0,0,1}
	\definecolor{CYAN}{cmyk}{1,0,0,0}
	\definecolor{MAGENTA}{cmyk}{0,1,0,0}
	\definecolor{YELLOW}{cmyk}{0,0,1,0}
}

\newcommand{\mrd}{\mathrm{d}}

\newcommand{\mr}[1]{\mathrm{#1}}

\definecolor{darkred}{rgb}{0.9,0.05,0.05}
\definecolor{darkblue}{rgb}{0.05,0.05,0.6}
\definecolor{darkgreen}{rgb}{0.05,0.6,0.05}
\definecolor{brightgreen}{rgb}{0.1,0.9,0.1}

\renewcommand*{\eqref}[1]{%
	\begingroup
	\hypersetup{
		linkcolor=darkblue,
		linkbordercolor=darkblue,
	}%
	\textcolor{darkblue}{(\ref{#1})}%
	\endgroup
}
\hypersetup{linkcolor=red,citecolor=darkgreen,urlcolor=darkred,colorlinks=true}

\begin{document}

\setlength{\skip\footins}{0.8cm}

\begin{titlepage}

\vspace{0.5cm}

\newcommand\blfootnote[1]{%
	\begingroup
	\renewcommand\thefootnote{}\footnote{#1}%
	\addtocounter{footnote}{-1}%
	\endgroup
}

\begin{center}
\renewcommand{\baselinestretch}{1.5}  %Line spacing
\setstretch{1.5}

{\fontsize{17pt}{12pt}\bf{Coupling Constants as Conserved Charges \\ in Black Hole Thermodynamics}}
 
\vspace{9mm}
\renewcommand{\baselinestretch}{1}  %Line spacing
\setstretch{1}

\centerline{\Large{Kamal Hajian$^{\dagger\ast}$\footnote{kamal.hajian@uni-oldenburg.de}\footnote{khajian@metu.edu.tr}} and \Large{Bayram Tekin$^{\ast}$\footnote{btekin@metu.edu.tr}}}
\vspace{4mm}
\normalsize
$^\dagger$\textit{Institute of Physics, University of Oldenburg, P.O.Box 2503, D-26111 Oldenburg, Germany}\\
$^\ast$\textit{Department of Physics, Middle East Technical University, 06800, Ankara, Turkey}
\vspace{5mm}

\begin{abstract}
\noindent In a generic theory of gravity coupled to matter fields, the Smarr formula for black holes does not work properly if the contributions of the coupling constants defining the theory are not incorporated. However, these couplings, such as the cosmological constant or the dimensionful parameters that appear in the Lagrangian, are fixed parameters defining the theory, and they cannot be varied. Here, we present a robust method, applicable to any covariant Lagrangian, that upgrades the role of the couplings from being constants in the theory to being free parameters of the solutions. To this end, for each one of the couplings in a theory, a pair of auxiliary scalar and gauge fields is introduced. The couplings are shown to be conserved charges of the global part of the implemented gauge symmetry. Besides, their conjugate chemical potentials are defined as the electric potential of the corresponding gauge fields on the black hole horizon. Using this method, we systematically extend the first law and the Smarr formula by coupling conserved charges and their conjugate potentials. The thermodynamics of a black hole solution in a quadratic gravity theory is given as an example. 
\end{abstract}
\end{center}

\let\newpage\relax
\end{titlepage}

\section{Introduction}

In the last decades, the cosmological constant $\Lambda$ \cite{Einstein:1917} has been the focus of astronomical observations \cite{Perlmutter:1998np,Riess:1998cb} as well as theoretical research such as the AdS/CFT correspondence \cite{Maldacena:1997re,Brown:1986nw} and the black hole thermodynamics \cite{Bardeen:1973gd,Bekenstein:1973ft,Hawking:1976rt} and chemistry \cite{Henneaux:1984ji,Henneaux:1989zc,Teitelboim:1985dp,Caldarelli:1999xj,Sekiwa:2006qj,Cvetic:2010jb,Dolan:2010ha,Dolan:2011xt,Mann_Kubiznak,Kastor:2009wy,Couch,Hyun:2017nkb,Hu:2018njr,Bokulic:2021dtz,Meessen:2022hcg,Xiao:2023lap}. Although a constant in the Lagrangian, it plays the role of pressure in black hole thermodynamics, which is traditionally a property of the solution. Despite this perplexing behavior, variation of the cosmological constant has to be included in the black hole first law to have a consistent Smarr formula \cite{Smarr:1972kt} (see a review in \cite{Mann_Kubiznak}). In addition, there are applications for the extension of black hole thermodynamics in the AdS/CFT as the central charge of the dual description \cite{Ahmed:2023snm}, as well as phase transitions \cite{Cong:2021fnf}, holographic complexity \cite{AlBalushi:2020rqe}, and weak cosmic censorship conjecture \cite{Harlow:2022ich}. 

 However, by fiat, $\Lambda$ is a constant in the Lagrangian and not a property of a particular solution (such as the mass or the charge or angular momentum of the solution). Thus, it is not clear how it can be varied to fit into the first law of black hole thermodynamics. Besides, its conjugate chemical potential lacks a firm geometrical prescription as a volume. 
The lack of a universal definition for the conjugate chemical potentials or the issue of constant parameter variation in the Lagrangian are two examples of how the literature, despite many excellent ideas and contributions, lacks a coherent and universal construction.
 
 As a remedy, the Lagrangian can be modified by an auxiliary gauge field\cite{Aurilia:1980xj, Duff:1980qv, Hawking:1984hk, Bousso:2000xa} such that the $\Lambda$ changes its role to be the conserved charge of the induced gauge symmetry and hence a parameter in the solution. In addition, its conjugate is defined naturally as the horizon gauge potential \cite{1, Hajian:2021hje}.       

The Smarr formula is democratic as far as the coupling constants other than $\Lambda$ are considered and cannot be satisfied unless other couplings are also included in the black hole thermodynamics. It suggests a general formulation that can capture all couplings as conserved charges in black hole physics. In this Letter, we provide a modification of the Lagrangian that can put the black hole chemistry with all couplings on firm ground. The main idea is the same as in  \cite{Aurilia:1980xj, Duff:1980qv, Hawking:1984hk, Bousso:2000xa,1, Hajian:2021hje}: the introduction of new symmetries. However, there are two main differences in the construction: (1) instead of a single gauge field
associated with the couplings, for each one of the couplings
(including the cosmological constant), a pair of fields (one gauge and one scalar field) is
implemented, and (2) the Lagrangian of the gauge  fields is no longer
quadratic but linear in the field strength. These differences are
crucial for a general method that works for an arbitrary coupling.

The Letter is organized as follows: In the next section, pairs of auxiliary fields are introduced in the Lagrangian, which allows one to interpret the coupling constants as parameters in the solutions. In Sec. \ref{sec-char}, we show that the couplings are conserved charges. Sec. \ref{sol-chem} is devoted to introducing conjugate chemical potentials. The first law of black hole thermodynamics and the Smarr formula are extended in Sec. \ref{sec-first}. In the last section, a clarifying example, which is a well-studied, 3D higher curvature gravity, is given to show the validity of our approach.

\section{Couplings as solution parameters}\label{sec-par}

In a $D$-dimensional space-time, we consider the action 
\begin{equation}\label{action}
I=\int \mrd^D x \sqrt{-g}\mathcal{L}, \qquad \mathcal{L}=\mathcal{L}_0-\sum_{i}\alpha_i\mathcal{L}_i,
\end{equation}
where $\mathcal{L}_0$  is Lagrangian that includes the kinetic term and comes with no coupling constant such as the Einstein-Hilbert term (Newton's constant is set to unity, or it multiplies the whole action); and $\mathcal{L}_i$'s are some number of other Lagrangian densities labeled by index $i$ that are coupled with the corresponding coupling constant $\alpha_i$. The densities can be, e.g., the (cosmological) constant term, higher curvatures, scalar tensor theories, Maxwell gauge fields, and in general, any covariant term that is built of the metric $g_{\mu\nu}$, curvature $R^{\alpha}_{\,\,\beta\mu\nu}$, covariant derivative $\nabla_\mu$, and other dynamical fields in the theory. The action \eqref{action} can be conventionally rewritten in terms of the volume $D$-form $\boldsymbol{\epsilon}$, i.e.,
\begin{equation}\label{action2}
\boldsymbol{\epsilon}=\dfrac{\sqrt{-g}}{D!}\epsilon_{\mu_1\dots\mu_D}\mrd x^{\mu_1}\wedge\dots\wedge\mrd x^{\mu_D}, \qquad I=\int\mathbf{L}=\int \left(\mathbf{L}_0-\sum_{i}\alpha_i \mathbf{L}_i \right),
\end{equation}
where we defined the $D$-forms as
\begin{equation}
\mathbf{L}\equiv \mathcal{L}\boldsymbol{\epsilon}, \qquad \mathbf{L}_0\equiv \mathcal{L}_0\boldsymbol{\epsilon}, \qquad \mathbf{L}_i\equiv \mathcal{L}_i\boldsymbol{\epsilon}
\end{equation}
with $\epsilon_{01\dots D-1}=+1$ for the Levi-Civita symbol. Variation of $\mathbf{L}$ with respect to all dynamical fields, collectively denoted by $\Phi(x)$ including the metric $g_{\mu\nu}$, yields
\begin{equation}\label{delL}
\delta \mathbf{L}=\boldsymbol{\mathrm{E}}^{\Phi}\delta \Phi + \mrd \boldsymbol{\Theta}
\end{equation}
in which the summation convention over the fields should be understood. Setting $\delta \mathbf{L}=0 $ yields the field equations $\boldsymbol{\mathrm{E}}^{\Phi}=0$ associated with each one of the fields in the set $\Phi$; appropriate boundary conditions must also be provided for the well-posedness of the problem. We note that $\boldsymbol{\mathrm{E}}^{\Phi}$ and $\boldsymbol{\Theta}$ are linear in terms of the Lagrangian components, namely,
\begin{equation}\label{eom original}
\boldsymbol{\mathrm{E}}^{\Phi}\delta\Phi=\left(\boldsymbol{\mathrm{E}}^{\Phi}_0-\sum_{i}\alpha_i \boldsymbol{\mathrm{E}}^{\Phi}_i \right)\delta\Phi, \qquad \boldsymbol{\Theta}=\boldsymbol{\Theta}_0-\sum_{i}\alpha_i \boldsymbol{\Theta}_i.
\end{equation}

It is possible to introduce pairs of auxiliary fields in the Lagrangian in order to promote $\alpha_i$'s to be free parameters in the solution, not the theory. Each pair, labeled also by the index $i$, is composed of one scalar field denoted by $\alpha_i(x)$ and one ($D-1$)-form gauge field $\boldsymbol{\mathrm{A}}_i$. The field strength $\boldsymbol{\mathrm{F}}_i\equiv\mrd \boldsymbol{\mathrm{A}}_i$ is a top-form that is invariant under the gauge transformation
\begin{equation}\label{gauge}
\boldsymbol{\mathrm{A}}_i \to \boldsymbol{\mathrm{A}}_i+\mrd \boldsymbol{\lambda}_i.
\end{equation}
Equipped with the auxiliary field pairs, a given action $I$ in \eqref{action2} can be modified to an extended action $\tilde{I}$ as
\begin{equation}\label{tildeI}
\tilde{I}=\int \tilde {\mathbf{L}}\equiv\int \Big(\mathbf{L}_0-\sum_{i}\alpha_i(x) (\mathbf{L}_i-\boldsymbol{\mathrm{F}}_i)\Big).
\end{equation}
This action is symmetric under the gauge transformation \eqref{gauge} since $\boldsymbol{\mathrm{F}}_i$ is gauge invariant by construction. Besides, it reproduces the dynamics of the fields in the original action \eqref{action2} on-shell. To this end, we denote the collection of the fields $\{\Phi,\alpha_i,\boldsymbol{\mathrm{A}}_i \}$ by $\tilde \Phi$. Then, the variation of \eqref{tildeI} followed by the standard integration by parts gives
\begin{equation}\label{tilde delA}
\delta \tilde{\mathbf{L}}={\boldsymbol{\mathrm{E}}}^{\tilde \Phi}\delta \tilde \Phi + \mrd \tilde{\boldsymbol{\Theta}},
\end{equation}
in which the modified variations read as 
\begin{align}
{\boldsymbol{\mathrm{E}}}^{\tilde \Phi}\delta \tilde \Phi = \boldsymbol{\mathrm{E}}^{\Phi}_0\ \delta \Phi-\sum_{i}&\left[\Big(\alpha_i(x) \boldsymbol{\mathrm{E}}^{\Phi}_i-\mrd \alpha_i(x)\frac{\partial \mathbf{L}_i}{\partial (\mrd \Phi)}\Big)\delta \Phi+ (\mathbf{L}_i-\boldsymbol{\mathrm{F}}_i)\delta \alpha_i(x)+\mrd \alpha_i(x)\delta \boldsymbol{\mathrm{A}}_i\right],\\
&\quad \tilde {\boldsymbol{\Theta}}=\boldsymbol{\Theta}_0-\sum_{i}\alpha_i(x) \boldsymbol{\Theta}_i+\sum_{i}\alpha_i(x) \delta \boldsymbol{\mathrm{A}}_i. 
\end{align}
For clarity, the $x$-dependency of the scalar fields $\alpha_i$ is shown explicitly. In order to find the field equations by the action principle $\delta \tilde {\mathbf{L}}=0$, the coefficients of $\delta \Phi$, $\delta \alpha_i(x)$, and $\delta \boldsymbol{\mathrm{A}}_i$ in \eqref{tilde delA} should vanish independently. The equations that arise from the last two terms yield the on-shell relations
\begin{equation}\label{eom new}
\boldsymbol{\mathrm{F}}_i=\mathbf{L}_i, \qquad \mrd \alpha_i(x)=0,
\end{equation}
respectively. The last equality above implies 
\begin{equation} \label{alpha free}
\alpha_i(x)= \text{const.},
\end{equation}
which means that $\alpha_i$'s are some \emph{free solution parameters} that are constant over space-time. Inserting this crucial result in the overall coefficient of $\delta\Phi$ in \eqref{tilde delA} the original equations of motion in \eqref{eom original} are recovered. 

The argument above shows that, as far as the dynamics of the fields $\Phi$ are concerned, one can use the Lagrangian $\mathbf{L}$ and $\tilde{\mathbf{L}}$ interchangeably. However, the main advantage of $\tilde{\mathbf{L}}$  is to promote the coupling constants $\alpha_i$'s in $\mathbf{L}$ to be free parameters of the solutions $\tilde{\Phi}$. This vantage point of view is important for the rest of this Letter, where we need the variations of $\alpha_i$'s as solution parameters to be considered in the first law of black hole thermodynamics. Interestingly, one can go further and show that not only $\alpha_i$'s can be considered as free parameters in the solutions, but they are also conserved charges which we discuss next.

\section{Coupling constants as conserved charges}\label{sec-char}

Motivated by the analysis in \cite{1, Hajian:2021hje} where the cosmological constant was reinterpreted as a conserved charge, this section is devoted to showing that in a theory that is described by the action $I$ in \eqref{action2} or equivalently by $\tilde I$ in \eqref{tildeI}, the coupling constants or the solution parameters $\alpha_i$ in \eqref{alpha free} are conserved charges associated with the {\it global } part of the gauge symmetries in \eqref{gauge}. For this purpose, the ``covariant phase space" formulation of charges, also known as the Iyer-Wald formulation \cite{Lee:1990gr,Wald:1993nt,Iyer:1994ys,Wald:1999wa} (initiated and followed in \cite{Ashtekar:1987hia,Ashtekar:1990gc,Crnkovic:1987at,Barnich:2001jy}) is apt. The formalism is reviewed, e.g., in \cite{Hajian:2015eha,Seraj:2016cym, Corichi:2016zac} and applied to various theories (e.g., see \cite{Nutku}). 

Let us first focus on the action $I$ in \eqref{action2}. In the covariant phase space method, the symplectic current $\boldsymbol{\omega}$ is defined by taking an exterior derivative of $\boldsymbol{\Theta}$ on the field configuration space, i.e., 
\begin{equation}\label{omega}
\boldsymbol{\omega}(\delta_1\Phi,\delta_2\Phi,\Phi)=\delta_1\mathbf{\Theta}(\delta_2\Phi,\Phi)-\delta_2\mathbf{\Theta}(\delta_1\Phi,\Phi)\,.
\end{equation}   
If the fields $\Phi$ and their variations $\delta \Phi$ satisfy the field equations and their linearized versions respectively, then the symplectic current is locally conserved, i.e., $\mrd \boldsymbol{\omega}=0$. As a result, it is possible to define the symplectic 2-form 
\begin{equation}\label{Omega}
\Omega(\delta_1\Phi,\delta_2\Phi,\Phi)\equiv \int_\Sigma \boldsymbol{\omega}(\delta_1\Phi,\delta_2\Phi,\Phi)\,, 
\end{equation}
which makes the field configuration space a phase space. The hypersurface $\Sigma$ is a Cauchy surface, and the result in \eqref{Omega} is independent of its choice by the conservation of $\boldsymbol{\omega}$ and the appropriate boundary conditions.

Having the symplectic form in hand, one can associate a charge variation $\delta H_\epsilon$ to a symmetry generator $\epsilon\equiv \{\xi^\mu, \lambda\}$ that is composed of a diffeomorphism $x^\mu\to x^\mu-\xi^\mu$ and some Maxwell (or Yang-Mills) gauge transformation $A_\mu\to A_\mu+\partial_\mu \lambda$. The charge variation is defined as $\delta H_\epsilon\equiv \delta_\epsilon\Phi\cdot\Omega$, which yields
\begin{align}\label{delta H xi}
\delta H_{\epsilon}(\Phi)&=\int_\Sigma \big(\delta\mathbf{\Theta}(\delta_\epsilon\Phi,\Phi)-\delta_\epsilon\mathbf{\Theta}(\delta\Phi,\Phi)\big) =  \int_\Sigma \big(\delta\mathbf{\Theta}(\delta_\epsilon\Phi,\Phi)-\mathcal{L}_\xi\mathbf{\Theta}(\delta\Phi,\Phi)\big),
\end{align}
where in the last equation the gauge invariance of $\boldsymbol{\Theta}$ is used. By the Cartan identity, $\mathcal{L}_\xi\mathbf{\Theta}=\xi\cdot\mrd \mathbf{\Theta}+\mrd(\xi\cdot \mathbf{\Theta})$, and the on-shell relation $\mrd \mathbf{\Theta}= \delta \mathbf{L}$ the charge variation in \eqref{delta H xi} is equal to
\begin{align}\label{surface term}
\int_\Sigma \Big(\delta(\mathbf{\Theta}(\delta_\epsilon\Phi,\Phi)-\xi\cdot\mathbf{L})-\mrd\big(\xi\cdot\mathbf{\Theta}(\delta\Phi,\Phi)\big)\Big)=\int_{\Sigma}\mrd\big(\delta \mathbf{Q}_\epsilon(\Phi)-\xi\cdot\mathbf{\Theta}(\delta\Phi,\Phi)\big).
\end{align} 
The last equality follows from the celebrated Noether current 
\begin{equation}\label{Noether J}
\mathbf{J}_\epsilon=\mathbf{\Theta}(\delta_\epsilon\Phi,\Phi)-\xi\cdot\mathbf{L}(\Phi), \qquad \mrd \mathbf{J}_\epsilon=0 \quad \Rightarrow\quad \mathbf{J}_\epsilon=\mrd \mathbf{Q}_\epsilon,
\end{equation}
in which the Poincar\'e lemma is used to introduce $\mathbf{Q}$ as the Noether charge density, and $\delta \mrd = \mrd \delta$ was also used. By Stokes' theorem, the right-hand side of \eqref{surface term} can be written as a surface integral:
\begin{align}\label{del H}
\delta H_{\epsilon}=\oint_{\partial\Sigma}\boldsymbol{k}_{\epsilon}\,, \qquad \qquad \boldsymbol{k}_{\epsilon}(\delta\Phi,\Phi)\equiv \delta \mathbf{Q}_\epsilon(\Phi)-\xi\cdot\mathbf{\Theta}(\delta\Phi,\Phi) \,.
\end{align} 
Notice that similar to the field equations and $\boldsymbol{\Theta}$ in \eqref{eom original}, $\boldsymbol{k}$ is also linear in the Lagrangian components of arbitrary action, e.g., for the action $I$ in \eqref{action2}:
\begin{equation}
\boldsymbol{k}=\boldsymbol{k}_0-\sum_{i}\alpha_i \boldsymbol{k}_i. 
\end{equation}

Now, we are ready to follow these steps verbatim, this time for the action $\tilde{I}$ in \eqref{tildeI}. Considering the additional gauge symmetry in \eqref{gauge}, the charge generator $\epsilon$ is extended to capture this feature, namely 
\begin{equation}
\tilde\epsilon\equiv \{\xi^\mu, \lambda,[\boldsymbol{\lambda}_i]\}
\end{equation}
for $i$ number of gauge generators $\{\boldsymbol{\lambda}_i\}$. Then, replacing $\mathbf{L}\to \tilde{\mathbf{L}}$ and $\boldsymbol{\Theta}\to \tilde{\boldsymbol{\Theta}}$ in \eqref{Noether J} and \eqref{del H} with multiple usage of the on-shell condition \eqref{alpha free} we find
\begin{equation}\label{k modify}
\delta H_{\tilde\epsilon}=\oint_{\partial\Sigma}\tilde{\boldsymbol{k}}_{\tilde\epsilon}\,, \qquad \qquad \tilde{\boldsymbol{k}}_{\tilde \epsilon}=\boldsymbol{k}_\epsilon+\sum_i\big(\xi\cdot\mathbf{A}_i\delta \alpha_i+\delta (\alpha_i\boldsymbol{\lambda}_i)\big).
\end{equation}
This relation can be used to calculate the charges of the diffeomorphism and gauge symmetries, and we will use it to find mass, angular momentum, entropy, and other black hole charges in the last section. However, here we focus on a very specific symmetry that is a global part of the gauge transformation \eqref{gauge}. The generator that we choose is proportional to  $ \{0, 0,[\boldsymbol{\lambda}_i]\}$ with only one nonzero gauge generator, calling it $\boldsymbol{\lambda}_j$, that satisfies $\mrd{\boldsymbol{\lambda}}_j=0$. The rest of gauge generators in $\{\boldsymbol{\lambda}_i\}$, i.e., all $\boldsymbol{\lambda}_i$ for $i\neq j$ vanish. To fix the normalization of the generator, we can divide it by the factor $|\boldsymbol{\lambda}_j|\equiv\oint_{\partial\Sigma}\boldsymbol{\lambda}_j$, which is a constant, i.e., independent of the arbitrarily chosen $\partial\Sigma$ as well as the $x^\mu$ (explained below).  So, we define
\begin{equation}\label{generators}
\hat{\boldsymbol{\lambda}}_j=\frac{\boldsymbol{\lambda}_j}{|\boldsymbol{\lambda}_j|}, \qquad \tilde\epsilon_j\equiv \{0, 0,\hat{\boldsymbol{\lambda}}_j\}.
\end{equation}
For such generators that are solely composed of the gauge transformations \eqref{gauge}, the only relevant part of the charge variation is the last term in \eqref{k modify}. So, 
\begin{equation}\label{charges}
\delta H_{\tilde\epsilon_j}=\oint_{\partial\Sigma}\delta (\alpha_j\hat{\boldsymbol{\lambda}}_j)=\delta \alpha_j,
\end{equation}
where the last equation is a result of the on-shell $x^\mu$-independency of $\alpha_i$, $\delta \hat{\boldsymbol{\lambda}}_j=0$, and the normalization convention in \eqref{generators}. This result is one of the main achievements of this work, because it shows that the coupling $\alpha_j$ is the conserved charge $H_{\tilde\epsilon_j}$,
\begin{equation}
H_{\tilde\epsilon_j}=\alpha_j.
\end{equation} 

In order to complete the argument, we clarify why $\oint_{\partial\Sigma}\boldsymbol{\lambda}_j$  is a constant if $\mrd \boldsymbol{\lambda}_j=0$. By the last term in \eqref{k modify}, the charge variation $\delta H$ for the generator $\boldsymbol{\lambda}_j$ is proportional to this surface integral. However, the transformation is an exact symmetry, i.e., $\delta _{\boldsymbol{\lambda}_j}\tilde{\Phi}=0$ whose symplectic current $\boldsymbol{\omega}$ vanishes. This feature makes the charges not only to be conserved, i.e., independent of some time coordinate, but also independent of the choice of the $\partial\Sigma$. The remaining $D-2$ coordinates that parametrize $\partial \Sigma$ are integrated out. So, no space-time and $\partial\Sigma$-dependency remain. 

\section{Conjugate chemical potentials for the coupling constants}\label{sol-chem}

The electric charge in Maxwell's electrodynamics is the charge of the global part of the $U(1)$ gauge symmetry $A\to A+\mrd \lambda$ with $\mrd \lambda=0$. For a black hole, its conjugate chemical potential is the electric potential, i.e., $\mathit{\Phi}_{_{\mr H}}\equiv\xi_{_{\mr H}} \cdot A$ calculated on the event horizon, in which $\xi_{_{\mr H}}$ is the horizon generating Killing vector field. Motivated by this potential, for the action $\tilde I$ in \eqref{tildeI} which has the gauge fields $\mathbf{A}_i$ with associated conserved charges $\alpha_i$, we can define their conjugate chemical potentials on the event horizon as 
\begin{equation}\label{conjugates}
\mathit{\Psi}^i_{_\mr{H}}\equiv\oint _{\mr{H}}\xi_{_{\mr{H}}}\cdot \mathbf{A}_i.
\end{equation}
Such a definition of chemical potential for a coupling constant was first introduced in the context of upgrading the cosmological constant to be conserved charge in \cite{1}. It has been proven that it reproduces the conjugate thermodynamic volume and has been examined for various examples \cite{Hajian:2021hje}. In the next section, we use the coupling constants as conserved charges \eqref{charges} and their conjugates \eqref{conjugates} to extend the first law of black hole thermodynamics.   

\section{Extension of the first law of black hole thermodynamics} \label{sec-first}

Let us consider a stationary black hole in the coordinates where the horizon generating Killing vector is given as $\xi_{_{\mr{H}}}=\partial_t+\mathit{\Omega}^n\partial_{\varphi^n}$ in which $n$ runs over the the axial isometries, and $\mathit{\Omega}^n$ are corresponding horizon angular velocities. In \cite{Wald:1993nt,Iyer:1994ys}, Iyer and Wald showed that entropy for nonextremal black holes is a conserved charge of this vector normalized by the Hawking temperature $T_{_{\mr H}}=\frac{\kappa_{_\mr{H}}}{2\pi}$ \cite{Hawking:1976rt}, where $\kappa_{_{\mr{H}}}$ is the surface gravity of the Killing horizon H. Later in \cite{Hajian:2013lna,Hajian:2014twa}, an infinite number of horizon-killing vectors whose charges are the entropy of extremal black holes were found in their near-horizon region. However, in the presence of electromagnetic gauge fields, integrability shows that the proposed Killing vectors (both for extremal and nonextremal) have missed a contribution from the gauge fields. In \cite{Hajian:2015xlp,Hajian:2022lgy} it was shown that to have integrable and gauge, as well as,  diffeomorphism-invariant conserved charges, their vector field generator should be augmented by some gauge transformations (the reader is invited to read \cite{Hajian:2016kxx,Ghodrati:2016vvf,Hajian:2016iyp,Hajian:2017mrf} for reviews and applications). Here, we focus on the action $\tilde{I}$ in \eqref{tildeI} that, in addition to some probable Maxwell field $A_\mu \mrd x^\mu$, also has the auxiliary gauge fields $\mathbf{A}_i$. Then, in appropriately chosen gauges, the generator of the integrable entropy is 
\begin{align}\label{entropy gen}
\tilde{\epsilon}_{_S}=\frac{2\pi}{\kappa_{_{\mr{H}}}}\{\xi_{_{\mr{H}}},-\mathit{\Phi}_{_{\mr H}},[-\mathit{\Psi}^i_{_{\mr H}}\hat{\boldsymbol{\lambda}}_i]\}.
\end{align}
Let us denote the symplectic symmetry generators of the mass $M$, angular momenta $J_n$, and electric charge $Q$ by $\tilde{\epsilon}_{_M}=\{\partial_t,0,[0]\}$, $\tilde{\epsilon}_{_{J_n}}=\{-\partial_{\varphi^n},0,[0]\}$ and $\tilde{\epsilon}_{_Q}=\{0,1,[0]\}$, respectively. They are assumed to be ``{exact}" symmetries, i.e., they satisfy $\delta_{\tilde\epsilon} \tilde{\Phi}=0$. Then, \eqref{entropy gen} reads
\begin{align}\label{local constraint}
\frac{\kappa_{_\mr{H}}}{2\pi}\tilde{\epsilon}_{_S}= \tilde{\epsilon}_{_M}-\mathit{\Omega}^n_{_\mr{H}} \tilde{\epsilon}_{_{J_n}}-\mathit{\Phi}_{_\mr{H}}\tilde{\epsilon}_{ _Q}-\sum_i\mathit{\Psi}^i_{_\mr{H}}\tilde{\epsilon}_i,
\end{align}
where definition of $\tilde{\epsilon}_i$ in \eqref{generators} was also used. 
However, by the linearity of charge variations ${\delta}H_{a\tilde{\epsilon}_1+b\tilde{\epsilon}_2}=a{\delta}H_{\tilde{\epsilon}_1}+b\,{\delta}H_{\tilde{\epsilon}_2}$ in \eqref{k modify}, the first law of black hole thermodynamics is derived as
\begin{equation}\label{first law}
T_{_\mr{H}}\delta S= \delta M - \mathit{\Omega}^n_{_\mr{H}}\delta J_n -\mathit{\Phi}_{_\mr{H}} \delta Q-\sum_i\mathit{\Psi}^i_{_\mr{H}}\delta \alpha_i.
\end{equation}
This method of proving the first law was first introduced in \cite{Hajian:2015xlp}, where the nonextended version of \eqref{first law} was derived directly from the local identity \eqref{local constraint}, i.e., without addressing the surfaces of integration on the horizon and at infinity (which is used in the Iyer-Wald proof of the first law). It works because the generators in \eqref{local constraint} are all ``{exact}" symmetries, and as a result, their charge variations in \eqref{k modify} are independent of the surface of integration \cite{Barnich:2003xg}.   

By dimensional analysis and scaling argument, the Smarr formula can be deduced  from the first law (see, e.g., \cite{Townsend:1997ku,Kastor:2009wy}). The same analysis for the extended first law yields

\begin{equation}\label{Smarr}
(D-3) M=(D-2)T_{_\text{H}}S+(D-2)\mathit{\Omega}^n_{_\text{H}} J_n+ (D-3)\mathit{\Phi}_{_\text{H}} Q +k^{(i)}\mathit{\Psi}^i_{_\text{H}} \alpha_i.
\end{equation} 
Details of the derivation can be found in Sec. 6 of Ref. \cite{Hajian:2021hje}. The factor $k^{(i)}$ is the scaling of the $\alpha_i$, i.e., if the length $l$ is scaled by a factor $z$ as $l \to z\times l$, then $\alpha_i \to z^{k^{(i)}}\times\alpha_i$. 

\section{Example: Thermodynamics of the rotating BTZ black hole in the New Massive Gravity}\label{sec-ex}

In \cite{Hajian:2021hje} some black hole solutions were studied of which the Smarr formula was not satisfied if the contribution of the couplings was not included. Here, we provide one of these as an example: that is the rotating Banados-Teitelboim-Zanelli
(BTZ) black hole solution of the new massive gravity theory \cite{Bergshoeff:2009hq} in the coordinates $x^\mu=(t,r,\varphi)$ \cite{Banados:1992wn,Clement:2009gq} 
\begin{equation}\label{example L}
\mathcal{L}=\frac{1}{16\pi}\left[R-2\Lambda-\beta\left(\frac{3}{8}R^2-R_{\mu\nu}R^{\mu\nu}\right )\right].
\end{equation}
The metric is given as 
\begin{align}
&\mrd s^2= -\Delta\mrd t^2 +\frac{\mrd r^2}{\Delta}+r^2(\mrd \varphi-\omega \mrd t)^2, \qquad \Delta\equiv -m+\frac{r^2}{\ell^2}+\frac{j^2}{4r^2}, \qquad \omega\equiv \frac{j}{2r^2},
\end{align}
for $\Lambda=\frac{-1}{\ell^2}+\frac{\beta}{4\ell^4}$, $\Lambda<0$, and $\beta>0$.  The black hole outer and inner horizons are at the radii $r_+$ and $r_-$, which satisfy $2r_\pm^2=\ell^2 (m\pm\sqrt{m^2-{j^2}/{\ell^2}})$, where $m$ and $j$ are free parameters of the solution. The thermodynamic properties of this solution are \cite{Clement:2009gq,Alkac:2012bz}
\begin{align}\label{Prop BTZ-NMG}
& M=(1+\frac{\beta}{2\ell^2})\frac{m}{8}, \qquad  J=(1+\frac{\beta}{2\ell^2})\frac{j}{8},\nonumber\\
&\mathit{\Omega}_\pm=\frac{r_\mp}{\ell r_\pm}, \qquad T_\pm=\frac{r_\pm^2-r_\mp^2}{2\pi \ell^2 r_\pm}, \qquad S_\pm=(1+\frac{\beta}{2\ell^2})\frac{\pi r_\pm}{2},
\end{align}
with the horizon-killing vectors $\xi_\pm=\partial_t+\mathit{\Omega}_\pm \partial_\varphi$. The mass and angular momentum follow from the general construction of conserved charges in higher derivative theories of gravity \cite{dt1,dt2}.

It is easy to check that the quantities in \eqref{Prop BTZ-NMG} do not satisfy the Smarr formula if the last term in \eqref{Smarr} is omitted. Examples such as this led us to the current attempt to rescue the Smarr formula. 
To remedy this issue, we can apply the procedure described above. The couplings $\Lambda$ and $\beta$ are promoted to the scalars $\Lambda(x)$ and $\beta(x)$, and their paired field strengths ${F}_{\!_\Lambda}(x)$ and ${F}_{\!_\beta}(x)$ [which are related to $\mathbf{F}_{\!_\Lambda}(x)$ and $\mathbf{F}_{\!_\beta}(x)$ in \eqref{tildeI} by a Hodge dual transformation] are implemented in the Lagrangian \eqref{example L}
\begin{equation}\label{example tilde L}
\tilde{\mathcal{L}}=\frac{1}{16\pi}\left(R-2\Lambda(x)\Big(1-{F}_{\!_\Lambda}(x)\Big)-\beta(x)\Big(\frac{3}{8}R^2-R_{\mu\nu}R^{\mu\nu}-{F}_{\!_\beta}(x)\Big)\right).
\end{equation}
It is clear that there is a conventional normalization in defining the couplings, e.g., instead of the $\Lambda$ one can consider $\frac{\Lambda}{8\pi}$ as the coupling. Nonetheless, as is expected, such a convention does not affect the physical thermodynamic laws because these factors are compensated in the conjugate potentials. 

By variation of the Lagrangian with respect to the new pairs of fields, equations of motion in \eqref{eom new} are derived that imply the following on-shell relations:
\begin{equation}
{F}_{\!_\Lambda}(x)=1, \qquad {F}_{\!_\beta}(x)=\frac{3}{8}R^2-R_{\mu\nu}R^{\mu\nu}, \quad 
\end{equation}
as well as the constancy of the $\Lambda$ and $\beta$. Therefore,
\begin{align}
&\mathbf{F}_{\!_\Lambda}(x)=\boldsymbol{\epsilon}=\sqrt{-g} \,\,\mrd t\wedge\mrd r\wedge\mrd \varphi=r \,\,\mrd t\wedge\mrd r\wedge\mrd \varphi, \\
&\mathbf{F}_{\!_\beta}(x)=\Big(\frac{3}{8}R^2-R_{\mu\nu}R^{\mu\nu}\Big)\boldsymbol{\epsilon}=\frac{3r}{2\ell^4}\,\,\mrd t\wedge\mrd r\wedge\mrd \varphi.
\end{align}
The gauge fields whose field strengths are calculated above are
\begin{align}
&\mathbf{A}_{_\Lambda}(x)=-\Big(\frac{r^2}{2}-\frac{\beta m\ell^2}{2\beta-4\ell^2}\Big)\mrd t\wedge \mrd \varphi,\\
&\mathbf{A}_{_\beta}(x)=-\Big(\frac{3r^2}{4\ell^4}-\frac{m(\beta-4\ell^2)}{4\ell^2(\beta-2\ell^2)}\Big)\mrd t\wedge \mrd \varphi.
\end{align}
Notice that the second term in each one of the parentheses is a pure gauge, which can be fixed by different methods, e.g., we have requested that the integrability of the black hole charges be respected by the new contribution of $\xi\cdot\mathbf{A}_i\delta \alpha_i$ in \eqref{k modify} ($\boldsymbol{k}_\epsilon$ can be found in \cite{Hajian:2021hje,Ghodrati:2016vvf}). Now, we can insert the gauge fields into \eqref{conjugates} to find the chemical potentials
\begin{equation}
\mathit{\Psi}^\Lambda_\pm=-\pi\Big({r_\pm^2}-\frac{\beta m\ell^2}{\beta-2\ell^2}\Big), \qquad \mathit{\Psi}^\beta_\pm=-\pi\Big(\frac{3r^2_\pm}{2\ell^4}-\frac{m(\beta-4\ell^2)}{2\ell^2(\beta-2\ell^2)}\Big).
\end{equation}
One can check that the first law and the Smarr formula are satisfied for each one of the horizons as
\begin{align}
&\delta M=T_\pm \delta S_\pm+\mathit{\Omega}_\pm \delta J+\mathit{\Psi}^\Lambda_\pm \delta(\frac{\Lambda}{8\pi})+\mathit{\Psi}^\beta_\pm \delta(\frac{\beta}{16\pi}),\\
&0=T_\pm  S_\pm+\mathit{\Omega}_\pm J-2\mathit{\Psi}^\Lambda_\pm (\frac{\Lambda}{8\pi})+2\mathit{\Psi}^\beta_\pm (\frac{\beta}{16\pi}),
\end{align} 
respectively. The numerical factors $\frac{1}{8\pi}$ and $\frac{1}{16\pi}$ in the coupling charges are conventional, and come from how   $\alpha_i$ and $\mathcal{L}_i$ are defined from the combination $\alpha_i\mathcal{L}_i$ in the Lagrangian \eqref{example tilde L}. However, independent of this convention, the relation
\begin{equation}
\mathit{\Psi}^\Lambda_\pm \delta(\frac{\Lambda}{8\pi})=V_\pm\delta P
\end{equation}
reproduces the volume-pressure term in black hole chemistry in which $V$ is the ``thermodynamic volume" introduced in \cite{Kastor:2009wy} (note the conventional minus signs that cancel each other in $V$ and $P$). This result is not accidental, and proof of it can be found in Sec. 2 of the Ref.\cite{Hajian:2021hje}.

\section{Conclusions}

When various dimensionful couplings enter a theory of gravity has matter fields, the cosmological constant, and higher derivative terms, the first casualty of black hole thermodynamics is the beautiful Smarr formula expressing the relation between the conserved charges. One has two options: accept that the Smarr formula is an accident of Einstein's gravity, or try to rescue it by upgrading the coupling constants of the theory to be conserved charges of the corresponding solution. But this requires a crucial step: first, the coupling constants are assumed to be space-time-dependent fields that are set to be constants as a consequence of the field equations. This can be done by the introduction of auxiliary Abelian gauge fields as described in this work. The vantage point presented here, which enlarges the theory by considering the coupling constants as space-time fields that take constant values as a result of the field equations, saves the Smarr formula. We have given a detailed account of this above and applied the new formalism to the rotating BTZ black hole in the new massive gravity (a quadratic theory much studied in the last decade).

The construction of all couplings and their contribution to black hole chemistry by the new symmetries is not a trivial task and may have far more consequences than the extended black hole thermodynamics.
Furthermore, interesting questions arise from considering couplings as conserved charges via the new Lagrangian in Eq. \eqref{tildeI}. The physical meaning of the corresponding pair of auxiliary fields and couplings as charges must be considered in the classical regime. Furthermore, it would be fascinating to investigate how the initial and boundary conditions determine them. It is also possible to investigate their dynamics and how they relate to the couplings renormalization group flows. Our understanding of the couplings in nature can be impacted at the quantum level by the quantization of the fields and its consequences for the couplings.

\vskip 1 cm

\noindent \textbf{Acknowledgements:} K.H. is thankful to Jutta Kunz for her support and to members of the HEP group at IPM for the useful discussions in the weekly meetings. This work has been supported by TUBITAK International Researchers Program No. 2221.

\end{document}